\documentclass{article}
\usepackage{spconf,amsmath,graphicx}
\usepackage{amsmath}
\usepackage{cite}
\usepackage{tabulary}
\usepackage{amssymb}
\usepackage{balance}

\usepackage[ruled,vlined]{algorithm2e}

\usepackage{color}

\title{EMG wrist-hand motion recognition system for real-time Embedded platform}
\name{Sumit Raurale, John McAllister, and Jesus Martinez del Rincon\thanks{Corresponding author's e-mail: sraurale01@qub.ac.uk}}
\address{The Institute of Electronics, Communications and Information Technology (ECIT), \\Queen's University of Belfast, U.K.}
\begin{document}
%
\maketitle
\begin{abstract}
Electromyography (EMG) signal analysis is a popular method for controlling prosthetic and gesture control equipment. For portable systems, such as prosthetic limbs, real-time low-power operation on embedded processors is critical, but to date there has been no record of how existing EMG analysis approaches support such deployments. This paper presents a novel approach to time-domain classification of multi-channel EMG signals harnessed from randomly-placed sensors according to the wrist-hand movements which caused their occurrence. It shows how, by employing a very small set of time-domain features, Kernel Fisher discriminant feature projection and Radial Bias Function neural network classifiers, nine wrist-hand movements can be detected with accuracy exceeding 99\% - surpassing the state-of-the-art on record. It also shows how, when deployed on ARM Cortex-A53, the processing time is not only sufficient to enable real-time processing but is also a factor 50 shorter than the leading time-frequency techniques on record. 
\end{abstract}


\section{Introduction}
\label{sec:intro}
Electromyography (EMG) has been used successfully as a control signal for a number of prosthesis and gesture control applications. In particular, skin-surface EMG signals recorded on the lower forearm can be used, via discrete analysis and discrimination methods, to identify a wide range of wrist-hand motions with accuracy approaching 100\% \cite{ref1,ref2,ref3}. 

These systems use various time-frequency analysis techniques which exploit the differing EMG frequency characteristics of the major arm muscles \cite{ref4,ref5,ref6}. A number of complex system architectures have been developed to accurately distinguish these signals according to wrist-hand motions which caused them \cite{ref2,ref3,ref6,ref7}. However in many applications - for example, prosthetic limbs - this capability needs to be deployed in a portable, battery-operated device. In all of the developments to date, so far as the authors are aware, no account has been made of computational complexity or real-time performance on an embedded device. 

This paper addresses these shortcomings. It presents a novel approach to identify hand open/close, wrist flexion/extension, wrist pronation/supination, wrist ulnar-flexion/ radial-flexion and relaxation using time-domain analysis of multi-channel EMG signals, where each channel has been acquired from a random position on the forearm. The proposed approach makes the following contributions:
\vspace{-0.15cm}
\begin{itemize}
\item It is shown how, by combining a small set of time-domain features with Kernel Fisher Discriminant analysis and Radial Basis Function (RBF) classification, the leading classification accuracy on record is enabled.
\vspace{-0.15cm}
\item It is shown how this approach can be realised on ARM Cortex-A53 to enable real-time identification of wrist-hand movements with a 5 ms latency. To the best of the authors' knowledge, this is the first real-time embedded implementation on record replying on randomly-placed sensing regime. 
\end{itemize}
\vspace{-0.15cm}

The remainder of this paper is as follows. Section 2 surveys the state-of-the-art in EMG wrist-hand motion classification techniques, before Section 3 presents our proposed approach and quantifies its detection accuracy and Section 4 its performance on ARM Cortex-A53.

\vspace{-0.15cm}
\section{Related work}
\label{sec:literature}
Skin-surface measurement of EMG signals is a widely adopted technique by which wrist-hand movements can be identified. In general, there are two main approaches to enabling this identification.

The first uses sensors located precisely over the major muscles on the lower arm \cite{ref2,ref3,ref7,ref8}. Given that the frequency content of the EMG emanating from different muscles exhibits different frequency characteristics, time-frequency domain analysis of the resulting EMG signals is usually employed in these cases. Different specific techniques have been employed, but in general all combine forms of feature extraction in order to represent the signal as a vector of features, followed by feature projection to reduce the vector's dimensionality and classification to discriminate each vector as belong to one of a fixed set of classes \cite{ref5,ref6,ref9}. The approach in \cite{ref2} uses the Delsys myoelectric system for EMG acquisition from four forearm muscles and employs Local Discriminant Basis with Wavelet Packet Transform for feature extraction, Principal Component Analysis (PCA) for feature projection and Multi-layer Perceptron (MLP) classification. In \cite{ref7} an Ottobock acquisition system acquires EMG from two forearm muscles to classify three hand motions using one-dimensional Local Binary Pattern feature extraction. MyoScan EMG sensors are used in \cite{ref10} to sense four forearm muscles classifying four hand motions using a variety of time/frequency features with PCA and Linear Discriminant Analysis (LDA) for feature projection and classification. Finally, \cite{ref3} uses eight electrodes, time domain analysis for feature extraction and Support Vector Machine classification. 

A second approach uses EMG sensors randomly placed on the skin's surface. Since these are not placed over specific muscles, they cannot avail of the variation in frequency characteristics between muscles, and employ time-domain analysis. In \cite{ICASSP18}, the Myo armband is used to acquire EMG signals from eight random skin-surface locations to classify nine wrist-hand motions with 95\% accuracy. It uses four time-domain features including 7th order auto-regression followed by LDA feature projection and MLP-based classification. 

All of these works focus on the classification accuracy. To the best of the authors' knowledge, none have considered the performance and cost of their implementation. In particular for portable applications such as prosthetics, it is critical that an embedded realisation can maintain real-time performance whilst minimising power consumption in order to maximise battery-life. For instance, in order to realise real-time operation for the experimental configuration in \cite{ICASSP18}, processing time needs to be lower than 128 ms. There is no record of whether this can be achieved in any existing system. In the majority of the cases in \cite{ref2,ref3,ref7,ref8}, bulky EMG acquisition and analysis equipment is required and normally signal analysis is performed on a PC. Neither of these are conducive to portability nor for portable or battery-powered operation. The real-time performance of existing solutions is benchmarked in Section \ref{sec:benchmarks}.

\section{Embedded EMG Analysis}
\label{sec:benchmarks}
To benchmark the real-time performance of existing EMG classification systems of the two types outlined in Section \ref{sec:literature}, optimised C++ derived via MATLAB's Embedded Coder \cite{matlabcoder} from the models used to compare the performance in \cite{ICASSP18} is considered. This is deployed on a 1.4 GHz ARM Cortex-A53 processor on a Raspberry Pi 3 B+ model. Table \ref{tableI} quotes the resulting processing times.
\begin{table}[htb]
\renewcommand{\arraystretch}{1.1}
\caption{Processing Time Benchmarks}
\centering
\begin{tabular}{r | cc}
\hline
\hline
     & \multicolumn{2}{c}{\textbf{Processing time (ms)}} \\
     \cline{2-3}
\textbf{System} & MATLAB & Embedded \\
\textbf{Configurations}  & Platform & Platform  \\
\hline
Time-freq system \cite{ref2} & 567.22 & 226.12 \\
\hline
Time-domain system \cite{ICASSP18} & 11.70 & 9.75 \\
\hline
\hline
\end{tabular}
\vspace{-0.15cm}
\label{tableI}
\end{table}

It is instructive to compare the execution time of the two realisations to the requirements of the Myo armband used in  \cite{ICASSP18}. These require a maximum processing time of 128 ms, the interval between EMG signal windows. Relative to this value, it is clear that the time-domain system in \cite{ICASSP18} comfortably enables real-time realisation, but the time-frequency system in \cite{ref2} does not. Indeed, the processing time of \cite{ICASSP18} is significantly lower than that in \cite{ref2}, by a factor of around 20. 

For portable applications real-time performance is critical, yet so is energy consumption. To mimimise energy consumption, absolute minimisation of workload is desired. Section \ref{sec:lowcomplex} explores alternatives to \cite{ICASSP18} with this objective.

\vspace{-0.15cm}
\section{Low Complexity EMG Analysis}
\label{sec:lowcomplex}
\vspace{-0.15cm}
The goal of our study is to develop a low-complexity EMG wrist-hand motion recognition system for portable applications. The system has four key components: EMG acquisition, feature extraction, feature projection and classification. 

\vspace{-0.20cm}
\subsection{EMG Acquisition}
\label{ssec:emgacq}
The Myo Armband has eight active electrodes which sample the skin-surface EMG sensors at a rate of 200 Hz per channel. The Myo armband is placed randomly on the subject's upper forearm so as to consider the maximum hand surface wherein the muscles are well sorted \cite{ref011}. All possible wrist-hand motions are considered \textit{viz.} opening and grasping of the fingers, flexion and extension of the wrist, pronation and supination of the wrist, radial and ulnar flexion of the wrist, and relaxation.

\vspace{-0.15cm}
\subsection{Feature Extraction}
\label{ssec:featext}
We propose to reduce the four features in \cite{ICASSP18} (including a 10-element vector feature) to three scalars: Integrated-EMG, natural logarithm of variance and Root Sum Square.

\vspace{-0.15cm}
\subsubsection{Integrated EMG (IEMG)}
\label{sssec:iemg}
The EMG signal shows change in potential levels from muscle contraction than the stationary state. The IEMG feature is used to detect the muscle activation condition as an onset index by illustrating the firing point within activation of muscle contraction levels of the EMG signal sequence. The IEMG of an $N$-element discrete signal sequence $x\left[t\right]$ is given by \cite{ref13}:
\begin{eqnarray}
\mathbf{f}_{IEMG} = \sum_{t=1}^N \mid{x\left[t\right]}\mid
\label{equ1}
\end{eqnarray}

\subsubsection{Natural log of Variance (lnVAR)}
\label{sssec:lvar}
Natural log of variance is used to estimate the power density of the EMG by characterising the differences in magnitude of the variance whilst amplifying relatively small absolute values. It is defined as \cite{ref13}:
\begin{eqnarray}
\mathbf{f}_{\ln VAR} = \ln \Bigg[ \dfrac{1}{N-1}\sum_{t=1}^{N}\mid x[t]-\mu \mid ^{2} \Bigg]
\label{equ2}
\end{eqnarray}
where $\mu$ is the mean of sample EMG signal $x$.

\subsubsection{Root Sum Square (RSS)}
\label{sssec:rss}
RSS represents the non-fatiguing in the muscle contraction levels while change in movement \cite{Ref14} occurs, allowing muscle strength variations to be detected. It is expressed as \cite{Ref14}:
\begin{eqnarray}
\mathbf{f}_{RSS} = \sqrt{\sum_{t=1}^{N} x[t]^{2}}
\label{equ3}
\end{eqnarray}

For the purposes of projection and classification, these features for each EMG across 8-channels are gather into a compound vector $\mathbf{f}_{feature} \in \mathbb{R}^{8}$, with the composite feature vector $\mathbf{f} \in \mathbb{R}^{24}$ as concatenation of the results, i.e.
\begin{eqnarray}
\mathbf{f} = [\mathbf{f}_{IEMG}, \mathbf{f}_{\ln VAR}, \mathbf{f}_{RSS}]
\label{equ4}
\end{eqnarray}

\subsection{Feature Projection}
\label{ssec:nldaproj}
Feature projection is used to remove any dimensional redundancy in the feature vector data whilst maximizing class separability for the subsequent classifier \cite{ref2,ref8,ref9}. In most approaches, LDA is used \cite{ref2,ref8,ref9}. However, if less well-differentiated features are input to LDA, linear discrimination leads to poorer between-class separation in the projected feature space \cite{ref15}. To avoid this problem, the use of non-linear Kernel Fisher Discriminant (KFD) analysis \cite{ref15} is proposed to project onto an eight-dimensional subspace from the input feature vector $\mathbf{f}$.

\subsection{Classification}
\label{ssec:RBFNN}
For classification, a Radial Basis Function (RBF) Neural Network is used including a singe hidden layer of fourteen RBF nodes which calculates the outcome of the basis functions, along with the output layer's nodes providing a linear combination of the basis functions. The nodes' selection criterion was based on the convergence of the learning error from different combinations. In the feed-forward RBF network, each node in hidden layer are entirely connected to the linear output units. The output nodes ($\Psi_{j}$) form a linear combination of the basis functions \cite{ref11} computed by each hidden layer nodes as shown in \eqref{equ4}.
\begin{eqnarray}
\Psi_{j}(\mathbf{s}) = K \bigg( \dfrac{\parallel \mathbf{s} - \mathbf{e}_{j} \parallel}{\sigma_{j}^{2}} \bigg)
\label{equ4}
\end{eqnarray}
The node output $\Psi_{j}$ is obtained by calculating the adjacency of the input to an $n$-dimensional vector $\mathbf{e}_{j}$ associated with the $j^{th}$ hidden unit; where $K$ is a positive radially symmetric function with a unique maximum at its centre $\mathbf{e}_{j}$ and which drops off rapidly to zero away from the centre.

A function $f:R^{n} \rightarrow R^{1}$ is approximated with an RBF network where the input scatter vector $\mathbf{s}\in R^{n}$ is an input, $\Psi(\mathbf{s},\textbf{e}_{j},\sigma_{j})$ is the $j$th function with centre $\mathbf{e}_{j}\in R^{n}$, width $\sigma_{j}$, and $\mathbf{w}$=$(w_{1},w_{2},...,w_{M})$ $\in R^{M}$ is the vector of linear output weights and $M$ is the number of basis functions used. The $M$ centres $\mathbf{e}_{j}\in R^{n}$ are concatenated into $\mathbf{e}$=$(e_{1},e_{2},...,e_{M}) \in R^{nM}$ and the widths to get $\sigma=(\sigma_{1},\sigma_{2},...,\sigma_{M})$ $\in R^{M}$. The output of the network for $\mathbf{s}\in R^{n}$ and $\sigma\in R^{M}$ is shown in \eqref{equ5}.
\vspace{-0.05cm}
\begin{eqnarray}
F(\mathbf{s},\mathbf{e},\sigma,\mathbf{w}) = \sum_{j=1}^{M} w_{j} \Psi(\mathbf{s},\mathbf{e}_{j},\sigma_{j})
\label{equ5}
\end{eqnarray}

\begin{figure*}
	\centering
	\includegraphics[width=0.95\linewidth]{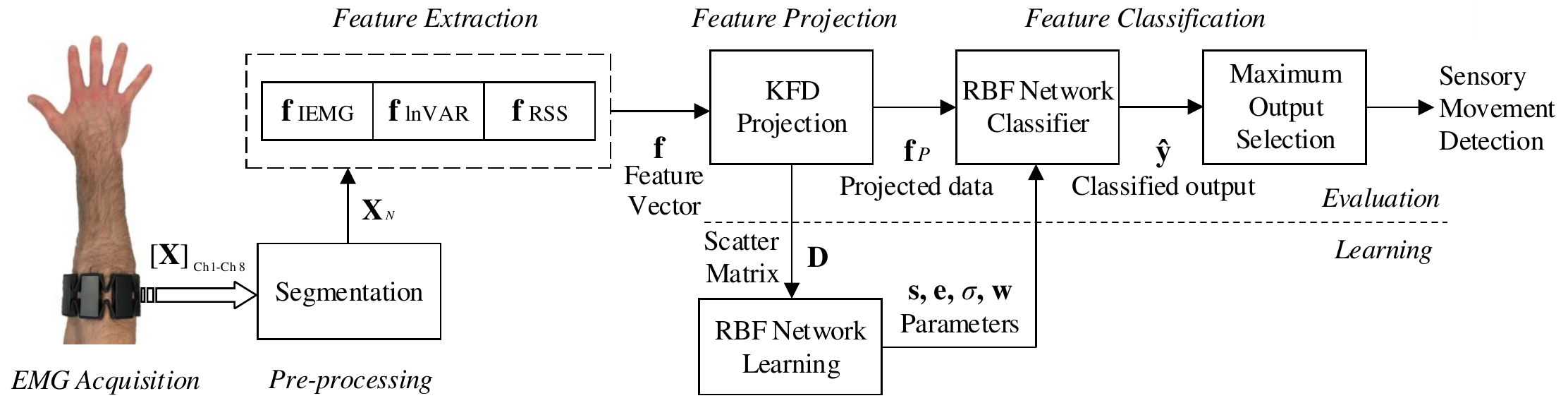}
	\caption{Proposed EMG wrist-hand motion recognition system}
    \label{EMGsys}
\end{figure*}

In evaluation phase, the maximum node value from output layer of the RBF network is selected as the recognized motion from a given KFD feature vector. The full architecture of the classification system is illustrated in Fig. \ref{EMGsys}.

\section{Experimental Results}
\label{sec:majhead}
The classification accuracy of the proposed system is evaluated by application to EMG recordings from ten normal subjects (seven males and three females, 26 $\pm$ 4 years). Twenty sessions were conducted for each motion performed for a 5 sec duration. To recognize a steady-state motion, a moving window scheme is applied with 250 ms windows separated by 125 classified in turn. 

The system is trained on data from ten sessions of ten subjects each. From each window is derived a feature vector $\mathbf{f}\in \mathbb{R}^{H}$ where $H$ is number of feature coefficients (24 as evaluated from Section \ref{ssec:featext}). The features derived from each window are combined into a feature matrix $\mathbf{G}\in \mathbb{R}^{J \times H}$, where $J=7200$ is the number of windows in the training data set. KFD derives from $\mathbf{G}$ and its associated class label vector a scatter matrix $\mathbf{D}\in \mathbb{R}^{J \times p}$ ($p=8$) as well as an eigenvector matrix $\mathbf{E}\in \mathbb{R}^{p \times H}$ and mean vector $\overline{\mathbf{f}}\in \mathbb{R}^{H}$. The eigenvector matrix $\mathbf{E}$ and mean vector $\overline{\mathbf{f}}$ are used in the evaluation phase while the scatter matrix $\mathbf{D}$ with its respective class label vector is used to train the RBF network producing trained network model parameters.

 Algorithm \ref{Algorithm} shows the steps taken in the evaluation phase to translate the segmented input signal $\mathbf{X}_{N}\in \mathbb{R}^{N \times c}$ which contains an $N=256$ window for each of the $c=8$ EMG channels, $\mathbf{E}$ and $\overline{\mathbf{f}}$ into an output vector $\widehat{\mathbf{y}}$ indicating the movement identified. Initially, each of the three feature vectors $\mathbf{f}_{IEMG}\in \mathbb{R}^{c}$, $\mathbf{f}_{\ln VAR}\in \mathbb{R}^{c}$, $\mathbf{f}_{RSS}\in \mathbb{R}^{c}$, are derived from $\mathbf{X}_{N}$ (line 1) before composition into $\mathbf{f}$ (line 2). This is then projected onto $\mathbf{f}_{P}\in \mathbb{R}^{p}$ by combining $\mathbf{f}$, $\overline{\mathbf{f}}$ and $\mathbf{E}$ (line 3) before evaluating the output vector $\widehat{\mathbf{y}}$ from trained RBF network classifier parameters.
 \begin{algorithm}[h]
 \KwIn{$\mathbf{X}_{N}$, $\overline{\mathbf{f}}$, $\mathbf{E}$}
 \KwOut{$\widehat{\mathbf{y}}$}
 \nl Analyse $\mathbf{f}_{IEMG}$, $\mathbf{f}_{\ln VAR}$, $\mathbf{f}_{RSS}$ from $\mathbf{X}_{N}$\;
 \nl Compute Feature vector $\mathbf{f} = [\mathbf{f}_{IEMG}, \mathbf{f}_{\ln VAR}, \mathbf{f}_{RSS}]$\;
 \nl Analyse project class vector $\mathbf{f}_{P}=(\mathbf{f}-\overline{\mathbf{f}})\times \mathbf{E}^{T}$\;
 \nl Evaluate output vector $\widehat{\mathbf{y}} \in \mathbb{R}^{v}$ as $v$-classes from trained RBF network model with $\mathbf{f}_{P}$\;
 \caption{\bf Proposed EMG wrist-hand motion recognition system in evaluation phase}
 \label{Algorithm}
 \end{algorithm}
 
The classification accuracy of the proposed system is compared with the approaches in \cite{ref2} and \cite{ICASSP18} for nine wrist-hand motions. The results are reported in Table \ref{classAccuracy}.
\begin{table}[!h]
\centering
\caption{Classification Accuracy}
\begin{tabular}{lccc}
\hline
& \multicolumn{3}{c}{{\small \textbf{Classification Accuracy (\%)}}}\\
\cline{2-4}
\vspace{-0.05cm}
{\small \textbf{Motion}} & {\small Time-freq}& {\small Time-domain} & {\small Proposed}\\
& {\small system}\cite{ref2} & {\small system}\cite{ICASSP18} & {\small system}\\
\hline
{\small Hand open} & {\small 76.00} & {\small 97.50} &  {\small 98.50} \\
{\small Hand close} & {\small 80.50} & {\small 97.70} &  {\small 98.80} \\
{\small Wrist flexion} & {\small 77.30} & {\small 97.60} &  {\small 99.20} \\
{\small Wrist extension} & {\small 70.50} & {\small 95.10} & {\small 99.10}  \\
{\small Wrist pronation} & {\small 61.90} & {\small 90.80} & {\small 98.00}  \\
{\small Wrist supination} & {\small 72.40} & {\small 94.90} &  {\small 99.20} \\
{\small Wrist ulnar flexion} & {\small 64.70} & {\small 91.30} & {\small 98.90}  \\
{\small Wrist radial flexion} & {\small 75.10} & {\small 96.80} &  {\small 99.20} \\
{\small Relaxation} & {\small 76.80} & {\small 98.40} & {\small 99.40}  \\
\hline
{\small \textbf{Total}} & {\small \textbf{72.80}} & {\small \textbf{95.57}} & {\small \textbf{99.03}}\\
\hline
\end{tabular}
\label{classAccuracy}
\end{table}

As reported, the average classification accuracy of 99.03\% is beyond not-only the time-frequency approach in \cite{ref2} but also the leading state-of-the-art time-domain approach in \cite{ICASSP18}. Metrics quantifying the real-time performance of the algorithm on both a 3.6 GHz Core i7 workstation PC running MATLAB R2016a platform and the same 1.4 GHz ARM Cortex-A53 on the Raspberry Pi 3 B+ platform used in Section \ref{sec:benchmarks} are quoted in Table \ref{tableIX}.
\begin{table}[htb]
\renewcommand{\arraystretch}{1.1}
\caption{Processing Time Metrics}
\centering
\begin{tabular}{r | cc}
\hline
\hline
     & \multicolumn{2}{c}{\textbf{Processing time (ms)}} \\
     \cline{2-3}
\textbf{System} & MATLAB & Embedded \\
\textbf{Configurations}  & Platform & Platform  \\
\hline
Time-freq system \cite{ref2} & 567.22 & 226.12 \\
\hline
Time-domain system \cite{ICASSP18} & 11.70 & 9.75 \\
\hline
Proposed system & 8.80 & 4.50 \\
\hline
\hline
\end{tabular}
\label{tableIX}
\end{table}

\vspace{0.1cm}
The processing time of the proposed pattern recognition system is 4.50 ms on the ARM Cortex-A53 - a figure significantly within both the required 128 ms processing time between EMG windows, and the previous lowest processing time recorded for the approach in \cite{ICASSP18}. This is significant. The dynamic power consumption of synchronous digital logic increases with the square of the frequency of the governing clock and hence if any way can be found to reduce this frequency then sharp reductions in power consumption can potentially result. Given the large differential between the processing time observed and that required for real-time operation (a factor of more than 25), very large reductions in power consumption could be realised via frequency scaling.

\section{Conclusion}
\label{sec:print}
A novel time-domain approach which determines wrist-hand motions from which EMG readings at the random skins surface originated is presented. By employing a very simple set of time-domains features to represent the activity on each of eight channels of EMG data alongside non-linear KFD-based feature projection and RBF-based classification, nine wrist-hand movements can be classified with accuracy exceeding 99\%. This is beyond the state-of-the-art for the same EMG acquisition regime. Furthermore, it is shown that the processing time on an embedded ARM Cortex-A53 processors is significantly below that required for real-time operation, indicating the potential for this approach to be realised on a low-power platform for portable, battery-operated applications.



\bibliographystyle{IEEEbib}
\balance
\bibliography{strings}

\end{document}